\documentclass{rmaa}

\usepackage{psfrag,color}
\usepackage[latin1]{inputenc}
\usepackage{bm}
\usepackage{amsmath}  
\usepackage{mathtools}
\usepackage{graphicx}

\title{Dynamics of clusters of galaxies with extended $f(\chi)$ gravity} 

\author{
  Tula Bernal,\altaffilmark{1} 
  Oliver L\'opez-Corona,\altaffilmark{2,3}
  and Sergio Mendoza\altaffilmark{2}}

\altaffiltext{1}{Universidad Aut\'onoma Chapingo}

\altaffiltext{2}{Instituto de Astronom\'{\i}a, Universidad Nacional Aut\'onoma de M\'exico.}
\altaffiltext{3}{Current address: Centro de Ciencias de la Complejidad, Universidad Nacional Aut\'onoma de M\'exico.}
  
\fulladdresses{
\item Tula Bernal: \'Area de F\'isica, Departamento de Preparatoria Agr\'icola, Universidad Aut\'onoma Chapingo, km.~38.5 Carretera M\'exico-Texcoco, 56230, Texcoco, Estado de M\'exico, M\'exico
(ac13341@chapingo.mx)
\item Oliver L\'opez-Corona: Instituto de Astronom\'{\i}a, Universidad Nacional Aut\'onoma de M\'exico, AP 70-264, Ciudad de M\'exico 04510, M\'exico. Current address:C\'atedras CONACyT Comisi\'on Nacional para el Conocimiento y Uso de la Biodiversidad (CONABIO), Liga Perif\'erico-Insurgentes Sur No. 4903, Parques del Pedregal, 14010 Tlalpan, CDMX, M\'exico. Red Ambiente y Sustentabilidad, Instituto de Ecolog\'ia A.C. de M\'exico (INECOL). Centro de Ciencias de la Complejidad (C3),UNAM.
(olopez@conacyt.mx)
\item Sergio Mendoza: Instituto de Astronom\'{\i}a, Universidad Nacional Aut\'onoma de M\'exico, AP 70-264, Ciudad de M\'exico 04510, M\'exico
    (sergio@astro.unam.mx)
}

\shortauthor{Bernal et al.}
\shorttitle{Clusters of galaxies with $f(\chi)$ gravity}

\SetVolume{00} \SetFirstPage{000} \SetYear{2019}

\ReceivedDate{Year Month Day} \AcceptedDate{Year Month Day} 

\resumen{En este art\'iculo presentamos los resultados del an\'alisis de perturbaciones al cuarto orden de la teor\'ia m\'etrica de gravedad $f(\chi) = \chi^{3/2}$, donde $\chi$ es un escalar de Ricci adimensional. Dicho modelo corresponde a una teor\'ia m\'etrica $f(R)$ en la que la masa del sistema est\'a incluida en la acci\'on del campo gravitacional.
En trabajos previos hemos mostrado que, hasta el segundo orden de perturbaciones, la teor\'ia reproduce las curvas de rotaci\'on planas de galaxias y los detalles de lentes gravitacionales en galaxias, grupos y c\'umulos de galaxias. Aqu\'i, dejando fijos los resultados de nuestros trabajos anteriores, mostramos que la teor\'ia reproduce las masas din\'amicas de 12 c\'umulos de galaxias del Observatorio \textit{Chandra} de rayos X, sin materia oscura, a partir de los coeficientes m\'etricos al cuarto orden de aproximaci\'on. En este sentido, calculamos la primera correcci\'on relativista de la teor\'ia m\'etrica $f(\chi)$ y la aplicamos para ajustar las masas din\'amicas de los c\'umulos de galaxias.}

\abstract{In this article, we present the results of a fourth order
perturbation analysis of the metric theory of gravity $f(\chi) =
\chi^{3/2}$, with $\chi$ a suitable dimensionless Ricci scalar. Such model
corresponds to a specific $f(R)$ metric theory of gravity, where the mass of
the system is included into the gravitational field's action.
In previous
works we have shown that, up to the second order in perturbations, this
theory reproduces flat rotation curves of galaxies and the details
of the gravitational lensing in individual, groups and clusters of
galaxies. Here, leaving fixed the results from our previous works, we show that the theory reproduces the dynamical masses of 12 \textit{Chandra} X-ray galaxy clusters, without the need of dark matter,
through the metric coefficients up to the fourth order of approximation.  In this sense, we calculate
the first relativistic correction of the $f(\chi)$ metric theory and apply it
to fit the dynamical masses of the clusters of galaxies.}


\begin{document}

\maketitle

\section{Introduction}
\label{introduction}

  From recent observations
of the European space mission {\it Planck}\rm, the contribution of the
baryonic matter to the total matter--energy density of the
Universe was inferred to be only $\sim5\%$, while the dark sector constitutes
$\sim95\%$, of which $\sim26\%$ is dark matter (DM) and $\sim 69\%$ is dark
energy (DE) or a positive cosmological constant $\Lambda$ \citep{Planck:2016}.
 The observations of type Ia supernovae, the anisotropies observed in the cosmic
microwave background, the acoustic oscillations in the baryonic matter, the
power--law spectrum of galaxies, among others,
represent strong evidences for the standard cosmological model, the
so--called $\Lambda$CDM {\it concordance model}\rm. 

 The DM component was postulated in order to explain the observed
rotation curves of spiral galaxies, as well as the mass-to-light ratios in
giant galaxies and clusters of galaxies \citep{Zwicky:1933,Zwicky:1937,Smith:1936,Rubin:1983}, the observed gravitational lenses and
the structure formation in the early Universe, among other astrophysical and
cosmological phenomena \citep[see e.g.][]{Bertone:2004,Bennett:2013}. On the other hand, the DE or $\Lambda$ has been postulated in order to explain the accelerated expansion of
the Universe \citep{SNIa2}.  
 The $\Lambda$CDM model adjusts quite well for most of these observations. 
However, direct or indirect searches of candidates to DM have yielded null
results. In addition, the lack of any further evidence for DE opens
up the possibility to postulate that there are no dark entities in the 
Universe, but instead, the theory associated to these astrophysical and 
cosmological phenomena needs to be modified.  

 Current models of DM and DE are based on the assumption
that Newtonian gravity and Einstein's general relativity (GR) are valid at all
scales. However, their validity has only been verified with high-precision
for systems with mass to size ratios of the order or greater that those of the Solar System. In that
sense, it is conceivable that both, the accelerated expansion of the Universe
and the stronger gravitational force required in different systems, represent
a change in our understanding of the gravitational interactions. 

 Moreover, from the geometrical point of view, theories of modified gravity are viable
alternatives to solve the astrophysical and cosmological problems that DM and DE are trying to solve
\citep[see e.g.][]{schimming04,nojiri-odintsov10,capozziello-book,Nojiri:2017}. In this
sense, any theory of modified gravity that attempts to supplant the DM of the Universe, must account for two crucial observations: the
dynamics observed for massive particles and the observations of the
deflection of light for massless particles.

 The first non-relativistic modification, proposed to fit the rotation curves 
and Tully-Fisher relation observed in galaxies, was the Modified Newtonian Dynamics (MOND)
\citep{milgrom83a,milgrom83b}. Due to its phenomenological nature and
its success in reproducing the rotation curves of disc galaxies \citep[see][for a review]{famaey11}, it is understood that any
fundamental theory of modified gravity should adapt to it on galactic scales
at the low accelerations regime. However, from the study of groups and
clusters of galaxies it has been shown that, even in the deep MOND regime, a
dominant DM component is still required in these systems (60 to 80\% of
the dynamical or virial mass). The central
region of galaxy clusters could be explained with a halo of neutrinos with
$2 \ \mathrm{eV}$ mass (which is about the upper experimental limit),  but
on the scale of groups of galaxies, the central contribution cannot be
explained by a contribution of neutrinos with the same mass \citep{angus-clusters}. Moreover,
the Lagrangian formulation of MOND/AQUAL \citep{Bekenstein:1984}, is not able to reproduce the
observed gravitational lensing for different systems
\citep[see e.g.][]{takahashi07,natarajan08}, mainly because it is not a relativistic theory and, as such, it cannot explain gravitational
lensing and cosmological phenomena, which require a relativistic theory of
gravity.

  Through the years, there have been some attempts to find a correct relativistic
extension of MOND. The first one was proposed by
\citet{bekenstein04}, with a Tensor-Vector-Scalar (TeVeS) theory.
This approach presents some cumbersome mathematical complications and it
cannot reproduce crucial astrophysical phenomena
\citep[see e.g.][]{sakellariadou09}.  Later, \citet{Bertolami:2007} showed that for a particular generalization of the $f(R)$ metric theories, with $R$ the Ricci scalar, by coupling the $f(R)$ function with the Lagrangian density of matter $\mathcal{L}_m$, an extra-force arises, which in the weak field limit can be connected with MOND's acceleration and explain the Pioneer anomaly. Also, \citet{bernal11a} showed that from the weak field limit of a particular $f(\chi)$ metric theory, with $\chi$ a dimensionless Ricci scalar, it is possible to recover the MONDian behavior in the metric formalism, as explained below. Another relativistic version of MOND recovers, in an empirical way, the MONDian limit through
modifications to the energy-momentum tensor \citep{demir-karahan14}.
\citet{Barrientos:2016} obtained the MONDian acceleration from an $f(\chi)$ theory in the Palatini formalism. In \citet{Barrientos:2017}, MOND's acceleration was obtained from an $f(R)$ metric theory of gravity with torsion. And more recently, \citet{Barrientos:2018b} showed that MOND's acceleration can be obtained in the weak field limit of a metric $F(R,\mathcal{L}_m)$ theory with a curvature-matter coupling.

In this article we focus on the $f(\chi)$ theories in the metric formalism, proposed in \citet{bernal11a}, where the dimensionless Ricci scalar $\chi$ is constructed by introducing a fundamental constant of nature with dimensions of acceleration of order $\sim10^{-10}\,\text{m/s}^2$. Through the inclusion of the mass of the system into the gravitational field's action, the authors showed that the $f(\chi)=\chi^{3/2}$ metric theory is equivalent to the MONDian
description in the non-relativistic limit for some systems, e.g.~for those with spherical symmetry, but
with remarkable advantages.  From the second order
perturbation analysis, such metric theory accounts in detail for two observational facts. First, it is
possible to recover the phenomenology of flat rotation curves and the
baryonic Tully-Fisher relation of galaxies, i.e.~a MOND-like weak field limit. Second, this construction also reproduces the details of observations of gravitational lensing in individual, groups and clusters of
galaxies, without the need of any DM component \citep{mendoza13}. At the same second order also, the theory is coherent with a Parametrized Post-Newtonian description where the parameter $\gamma=1$ \citep{mendoza14}.

The $f(\chi)$ metric theories are an extension of the $f(R)$ metric gravity, that has been extensively studied as an alternative to DM and DE \citep[see e.g.][]{sotiriou10,DeFelice:2010LRR,capozziello-book,nojiri11}.
In cosmology, it has been shown that $f(R)$ metric theories can account for the accelerated expansion of the Universe, as well as for an inflationary era, e.g.~for $f(R)=R + \alpha R^2$ \citep{Starobinsky:1980}, where $\alpha$ is a coupling constant. Moreover, there are models for unification of DE and inflation, or DE and DM \citep[see e.g.][]{Nojiri:2008,Nojiri:2017}.

As shown in \citet{carranza12}, an $f(\chi)$ description of gravity can be understood as a particular $F(R,T)$ theory \citep{Harko:2011}, where the gravitational action is an arbitrary function of the Ricci scalar and the trace of the energy-momentum tensor $T$.
 Within the $F(R,T)$ description, \citet{Harko:2011} have shown that, through the choice of suitable $F(T)$ functions, it is possible to obtain arbitrary FLRW universes, and that the model is equivalent to have an effective cosmological constant.
For the particular $f(\chi)=\chi^{3/2}$ metric theory used here, \citet{carranza12} have shown that the model can fit data of type Ia supernovae with a dust FLRW Universe, showing that the accelerated expansion of the Universe at late times ($\chi\sim1$) could be explained by an extended theory of gravity deviating from GR at cosmological scales.

In general, both descriptions result in field equations that depend on the mass of the source, except for the particular case $f(\chi)=\chi$, where GR is recovered. This scenario presents even more richness than standard $f(R)$ theories, because of the matter-geometry coupling, since it is possible to reconstruct diverse cosmological evolution by choosing different functions of the trace of the stress-energy tensor. Further research on the cosmological implications of $f(\chi)$ theories must be done in order to approve or discard models trying to replace the DM, or the DM and the DE of the Universe.

 In the present work, we extend the perturbation analysis developed in \citet{mendoza13} for $f(\chi)=\chi^{3/2}$ in the metric formalism, up to the fourth
order of the theory in powers of $v/c$ (where $v$ is the velocity of the components of the
system and $c$ is the speed of light), and focus on applications to clusters of galaxies. As shown in \citet{sadeh-lahav2015} and
first hypothesized in \citet{wojtak2011}, there exist observational
relativistic effects of the velocity of galaxies at the edge of galaxy
clusters, showing a difference of the inferred background potential with the
galaxy's inferred potential.  With this motivation in mind, we have calculated the
fourth order relativistic corrections and shown that they can fit the
observations of the dynamical masses of 12 \textit{Chandra} X-ray clusters of galaxies from
\citet{vikhlinin06}.

  The article is organized as follows: In Section~\ref{weak-field-limit}, the
weak field limit for a static spherically symmetric metric of any theory of
gravity is established and we define the orders of perturbation to be used
throughout the article.  In Section~\ref{extended-gravity}, we show the
particular metric theory to be tested with
astronomical observations.  The results from the perturbation theory for the
vacuum field equations, up to the fourth order in perturbations, are presented. With these results, we obtain the gravitational
acceleration generated by a point-mass source and its generalization to
extended systems, particularly for applications to clusters of galaxies. In
Section~\ref{cluster-profiles}, we establish the calibration method to fit the free parameters from the metric coefficients, from the observations of the dynamical masses of 12 \textit{Chandra} X-ray
clusters of galaxies.  Finally, in Section~\ref{discussion}, we show the results and the discussion.

\section{Perturbations in spherical symmetry}
\label{weak-field-limit}

  In this Section, we define the relevant properties of the perturbation
analysis for applications to any relativistic theory of gravity. Einstein's summation convention over repeated indices is
used. Greek indices take values $0,1,2,3$ and Latin ones $1,2,3$. In
spherical coordinates  $ (x^0,x^1,x^2,x^3) = (ct,r,\theta,\varphi) $, with
$ t $ the time coordinate and $ r $ the
radial one, $ \theta $ and $ \varphi $ the polar and azimuthal
angles respectively;  the angular displacement is $ \mathrm{d} \Omega^2 :=
\mathrm{d}\theta^2 + \sin^2\theta \, \mathrm{d}\varphi^2 $.  We use a
$ (+,-,-,-) $ signature for the metric of the space-time.

  Let us consider a fixed point-mass $ M $ at the center of coordinates; in this case, the static, spherically symmetric metric $ g_{\mu\nu} $ is generated by the interval
\begin{equation}
  \mathrm{d}s^2 = g_{\mu\nu} \mathrm{d}x^\mu \mathrm{d} x^\nu  
    = g_{00} \, c^2 \mathrm{d}t^2 + g_{11}\mathrm{d}r^2
	-r^2 \mathrm{d} \Omega^2 ,
\label{metric}
\end{equation}
where, due to the symmetry of
the problem, the unknown functions $ g_{00} $ and $ g_{11} $ are functions
of the radial coordinate $ r $ only.

  The geodesic equations are given by
\begin{equation}
  \frac{ \mathrm{d}^2 x^\alpha }{ \mathrm{d} s^2 } +
\Gamma^\alpha_{\mu\nu} \frac{
    \mathrm{d} x^\mu}{ \mathrm{d} s } \frac{ \mathrm{d} x^\nu }{ 
    \mathrm{d} s } = 0 \, ,
\label{geodesic}
\end{equation}
where $\Gamma^\alpha_{\mu\nu}$ are the Christoffel symbols. In the
weak field limit, when the speed of light $c \to \infty$, $\mathrm{d}s =
c \ \mathrm{d}t$, and since the velocity $v \ll c$, then each component $v^i
\ll \mathrm{d}x^0 / \mathrm{d}t$, with $v^i := \left( \mathrm{d}r /
\mathrm{d}t,\ r \, \mathrm{d} \theta / \mathrm{d}t,\ r\sin \theta \,
\mathrm{d} \varphi / \mathrm{d}t \right)$. In this case, the radial component
of the geodesic equations~(\ref{geodesic}), for the interval~(\ref{metric}),
is given by
\begin{equation}
  \frac{1}{c^2} \frac{\mathrm{d}^2 r}{\mathrm{d} t^2} = 
   - \frac{1}{2} g^{11} g_{00,r} \, ,
\label{radial:geodesic}
\end{equation}
where the subscript $( \ )_{,r} := \mathrm{d} / \mathrm{d} r$
denotes the derivative with respect to the radial coordinate $r$.
In this limit, a particle bounded to a circular orbit around the mass $M$
experiences the radial acceleration given by
equation~(\ref{radial:geodesic}), such that
\begin{equation}
  |\bm{a}_c| = \frac{v^2}{r} =	\frac{c^2}{2} g^{11} g_{00,r}
\label{centrifugal}
\end{equation}
 for a circular or tangential velocity $v$. At this point, it is
important to note that the last equation is a general kinematic relation, and
does not introduce any particular assumption about the specific gravitational
theory. In other words, it is completely independent of the field equations
associated to the structure of space-time produced by the energy-momentum
tensor.

  In the weak field limit of the theory, the metric coefficients take the following form
\citep[see e.g.][]{landaufields}:
\begin{eqnarray}
  g_{00} &=& 1 + \frac{ 2 \phi }{ c^2 },  \qquad g_{11} = -1 +
	\frac{ 2 \psi }{ c^2 },  
\label{weakest-metric}
\end{eqnarray}
 for the Newtonian gravitational potential $\phi$ and an extra
gravitational potential $\psi$.
As extensively described in
\citet{will93,will06}, when working with the weak field limit of a relativistic
theory of gravity, the dynamics of
massive particles determines the functional form of the time-time component $g_{00}$ of the 
metric, while the deflection of light determines the form of the radial $g_{11}$.
At the weakest order of the theory, the
motion of material particles is described by the potential $ \phi $,
taking $\psi=0$ \citep{landaufields}. The
motion of relativistic massless particles is described by taking into
consideration not only the second order corrections to the potential
$ \phi $, but also the same order in perturbations of the potential $ \psi $
\citep{will93}. 

  For circular motion about a mass $ M $ in the weak field limit of the
theory, the equations of motion are obtained when the left-hand side of
equation~(\ref{radial:geodesic}) is of order $ v^2 / c^2 $ and when the
right-hand side is of order $ \phi / c^2 $.  Both are orders
$ \mathcal{O}(c^{-2}) $ of the theory, or simply $ \mathcal{O}(2) $.
When lower or higher order corrections of the theory are introduced, we use the
notation $ \mathcal{O}(n) $ for $ n=0,1,2,\ldots$ meaning
$ \mathcal{O}(c^0), \ \mathcal{O}(c^{-1}), \ \mathcal{O}(c^{-2}),\ldots $,
respectively.

  Now, the extended regions of clusters of galaxies need a huge amount of DM
to explain the observed velocity dispersions of stars and gas in
those systems.  At the outer regions, the velocity dispersions are typically of
order $10^{-4}-10^{-3}$ times the speed of light.  Hence, the Newtonian
physics given by an $\mathcal{O}(2)$ approximation might be extended to
post-Newtonian $\mathcal{O}(4)$ corrections or, equivalently in our model,
``post-MONDian'' physics.

  In order to test a gravitational theory through different astrophysical
observations (e.g. the motion of material particles, the bending of
light-massless particles, etc.), the metric tensor $ g_{\mu\nu} $ is
expanded about the flat Minkowski metric $\eta_{\mu\nu}$, for corrections
$h_{\mu \nu} \ll \eta_{\mu\nu}$, in the following way:
\begin{equation}
   g_{\mu\nu} = \eta_{\mu\nu} + h_{\mu\nu} .
\end{equation}
 The metric $g_{\mu\nu}$ is approximated up to second perturbation
order $\mathcal{O}(2)$ for the time and radial components and up to zeroth
order for the angular components, in accordance with the spherical symmetry of
the problem.  At this lowest perturbation order, \citet{mendoza13} found the
time $g_{00}^{(2)}$ and radial $g_{11}^{(2)}$ metric components, for
the $f(\chi) = \chi^{3/2}$ metric theory of gravity.  These metric values are necessary to
compare with the astrophysical observations of motion of material
particles and that of photons through gravitational lensing
\citep{will93,will06}.  In fact, through the observations of the rotation
curves of galaxies and the Tully-Fisher relation, and the details of the
gravitational lensing in individual, groups and clusters of galaxies,
\citet{mendoza13} fixed the unknown potentials $\phi$ and $\psi$ of
the theory.

  In this paper, we develop perturbations of the relativistic extended model
$f(\chi) = \chi^{3/2}$ up to the fourth order in the time-time metric component,
$g_{00}^{(4)}$, corresponding to the next order of approximation to describe
the motion of massive particles \citep{will93}. In this case, the metric
components can be written as
\begin{eqnarray}
	g_{00} &=& {}1 + {}g_{00}^{(2)} + {}g_{00}^{(4)}
	  + \mathcal{O}(6) , \label{metric-perturbed:g00} \\
	g_{11} &=& - 1 + {}g_{11}^{(2)} + \mathcal{O}(4) .
\label{metric-perturbed}
\end{eqnarray}
 In other words, the metric is written up to the fourth order in the
time component and up to the second order in the radial one.  The
contravariant metric components of the previous set of equations are given by
\begin{eqnarray}
	g^{00} &=& 1 - {}g_{00}^{(2)} - {}g_{00}^{(4)} + \mathcal{O}(6) , \label{metric:exp-upg00} \\
	g^{11} &=& - 1 - g_{11}^{(2)} +	\mathcal{O}(4) .
\label{metric:exp-up}
\end{eqnarray}

\section{Extended $f(\chi)$ metric theories}
\label{extended-gravity}

\subsection{Field equations}
\label{field-eqs}

  The $f(\chi)$ metric theories, proposed in \citet{bernal11a}, are
constructed through the inclusion of MOND's acceleration scale $a_0$
\citep{milgrom83b,milgrom83a} as a fundamental physical constant, that has been shown to
be of astrophysical and cosmological relevance \citep[see e.g.][]
{bernal11b,carranza12,mendoza11,mendoza12,hernandez10,hernandez12a,
hernandez12b,mendoza13,mendoza14,mendoza15}.

  The action $S_\mathrm{f}$ for metric theories of gravity, rewritten with correct dimensional quantities for a
mass $M$ generating the gravitational field, is given by
\citep{bernal11a}
\begin{equation}
   S_\mathrm{f}  = - \frac{ c^3 }{ 16 \pi G } \int{ \frac{f(\chi)}{L_M^2} \, \sqrt{-g}
     \, \mathrm{d}^4x} \, ,
\label{action}
\end{equation}
where $G$ represents Newton's gravitational constant, for any arbitrary function $ f(\chi) $ of the
dimensionless Ricci scalar $\chi$:
\begin{eqnarray}
  \chi &:=& L_M^2 R;
\label{chi} \\
	L_M &:=& \zeta \left( r_\textrm{g} l_M \right)^{1/2},  
\label{def:LM}
\end{eqnarray}
where $L_M$ is a length-scale depending on the gravitational radius $r_\mathrm{g}$ and the mass-length scale $l_M$ of the system, given by \citep{mendoza11}
\begin{equation}
	r_\mathrm{g} := \frac{GM}{c^2}, \qquad l_M:=\left( \frac{GM}{a_0}
	\right)^{1/2},
\label{defs}
\end{equation}
 with $a_0 = 1.2 \times 10^{-10} \, \mathrm{m} / \mathrm{s}^2 $ the MOND's acceleration constant
\citep[see e.g.][and references therein]{famaey11} and $\zeta$ is a coupling constant of order one calibrated
through astrophysical observations.

  The matter action takes its ordinary form
\begin{equation}
    S_\mathrm{m} = - \frac{ 1 }{ 2 c } \int{ {\cal L}_\mathrm{m} \, \sqrt{-g} \,
    \mathrm{d}^4x } \, ,
\label{matter-action}
\end{equation}
 with $ {\cal L}_\mathrm{m} $ the Lagrangian matter density of the
system.
 
  Equation~\eqref{action} is a particular case of a full gravity-field action
formulation in which the details of the mass distribution appear inside the
gravitational action through $L_M$, except for $f(\chi)=\chi$, where the Hilbert-Einstein action is recovered. For the particular case of spherical symmetry, the mass inside action~\eqref{action} becomes the mass of the central object generating the gravitational field. It is also expected that for systems with high degree of symmetry, the mass $M$ is related to the trace of the energy-momentum tensor $T$ through the standard definition
\begin{equation}
    M:=(4\pi/c^2) \int T r^2 \mathrm{d}r .
\end{equation}

In what follows, we work with $f(\chi)$ theories in the metric formalism. Note that a metric-affine formalism can also be taken into account \citep[see e.g.][]{Barrientos:2016}.
 
  Now, the null variation of the complete action, i.e. $ \delta
\left( S_\mathrm{f} + S_\mathrm{m} \right) = 0 $, with respect to the metric
tensor $ g_{\mu\nu} $, yields the following field equations:
\begin{eqnarray}
  & f^\prime(\chi) \chi_{\mu\nu}   -  \frac{1}{2} f(\chi)
     g_{\mu\nu} -  L_M^2 \left( \nabla_\mu \nabla_\nu -g_{\mu\nu}
   \Delta \right) f'(\chi) \nonumber \\
  & = \frac{ 8 \pi G L_M^2 }{ c^4} T_{\mu\nu},
\label{field:eqs}
\end{eqnarray}
 where the prime denotes the derivative with respect to the argument,
the Laplace-Beltrami operator is $\Delta := \nabla^\mu \nabla_\mu$ and the
energy-momentum tensor $ T_{\mu\nu} $ is defined through the standard
relation $ \delta \mathcal{L}_{\textrm m} = - (1/2c) T_{\alpha\beta} \delta
g^{\alpha\beta} $.  Also, in equation~(\ref{field:eqs}), the dimensionless
Ricci tensor is defined as
\begin{equation}
	\chi_{\mu\nu} := L_M^2 R_{\mu\nu} \, ,
\label{def:chi}
\end{equation}
 where  $R_{\mu\nu}$ is the standard Ricci tensor.
The trace of equations~\eqref{field:eqs} is given by
\begin{equation}
  f^\prime(\chi) \, \chi  - 2 f(\chi) + 3 L_M^2  \, \Delta  f^\prime(\chi) =
    \frac{ 8 \pi G L_M^2 }{ c^4} T \, ,
\label{trace}
\end{equation}
 where $ T := T^\alpha {}_\alpha $.

  \citet{bernal11a} and \citet{mendoza13} have shown that the function
$ f(\chi) $ must satisfy the following limits:
\begin{equation}
	f(\chi) = 
	\begin{cases}
          \chi, \qquad \mathrm{when } \ \chi \gg 1 \textrm{ (General relativity)}, \\
          \chi^{3/2}, \quad \mathrm{when } \ \chi \ll 1 \textrm{ (Relativistic MOND)},
        \end{cases}
\label{f-chi-real}
\end{equation}
 in order to recover Einstein's GR in the limit
$ \chi \gg 1 $ and a relativistic version of MOND in the regime
$ \chi \ll 1 $.

Now, a complete extended cosmological model without the introduction of any
DM and/or DE components should explain several cosmological
observations, e.g. the cosmic microwave background, large scale structure formation, baryonic acoustic oscillations, etc.
However, when  mass-energy to scale ratios reach sufficiently large
numbers, of the order or greater than the ones associated to the Solar
System, then GR must be the correct theory to describe them.
In this direction, \citet{mendoza12} has proposed a ``transition function'' between both regimes, GR and ``relativistic MOND'', to describe the complete cosmological evolution:
\begin{equation}
    f(\chi) = \chi^{3/2} \frac{1 \pm \chi^{1+p}}{1 \pm \chi^{3/2+p}} \rightarrow
    \begin{cases}
      \chi^{3/2}, \quad \text{for} \quad \chi \ll 1, \\
      \chi, \qquad \text{for} \quad \chi \gg 1.
    \end{cases}
    \label{transition}
\end{equation}
For this function, GR is recovered when $\chi \gg 1$ in the strong field regime and the relativistic version of MOND is recovered when $\chi \ll 1$ in the weak field limit. Some observations suggest an abrupt transition between both limits of function~\eqref{f-chi-real} \citep[see][]{mendoza13,Hernandez,mendoza15}, meaning that it might be possible to choose the following step function to describe the evolution of the Universe:
\begin{equation}
    f(\chi) = 
    \begin{cases}
      \chi^{3/2}, \quad \text{for} \quad \chi \leq 1, \\
      \chi, \qquad \text{for} \quad \chi \geq 1.
    \end{cases}
    \label{transition2}
\end{equation}
However, in this work we are interested in the regime where GR should be modified, which in our case corresponds to the relativistic regime of MOND, assuming that where GR works well there should not be a modification. Thus, in the following, we work in the limit $\chi \ll 1$ only.

Note that the ``transition functions'' \eqref{transition} and \eqref{transition2} converge to GR
at very early cosmic times, when inflation should dominate the behavior of
the Universe.  This can be thought of a correct limit by including an
inflaton field for the exponential expansion of the Universe, or one can
think that at these very early epochs the $f(\chi)$ function should be
proportional to the square of the Ricci scalar in such a way that a
\citet{Starobinsky:1980} exponential expansion is reached \citep[see also][]{Nojiri:2017}.

\subsection{Relativistic MOND ($\chi \ll 1$)}

  For the case $\chi \ll 1$, the first two terms on the left-hand
side of equation~\eqref{trace} are much smaller than the third one, i.e.~$f^\prime(\chi) \, \chi  - 2 f(\chi) \ll  3 L_M^2  \, \Delta  f^\prime(\chi)$, 
at all orders of approximation \citep{bernal11a}. This fact means that the 
trace~\eqref{trace} can be written as
\begin{equation}
	3 L_M^2 \Delta f^\prime(\chi) = \frac{8 \pi G L_M^2}{c^4} T .
\label{delta:chi}
\end{equation}
  For the field produced by a point mass $M$, the right-hand side of last
equation~\eqref{delta:chi} is null far from the source and so, the last
relation in vacuum at all perturbation orders can be rewritten as
\begin{equation}
	\Delta f^\prime(\chi) = 0 .
\label{null:delta}
\end{equation}

  Now, as a simple case of study, we assume a power-law form for the function
$ f(\chi)$:
\begin{equation}
	f(\chi) = \chi^b ,
\label{power-law}
\end{equation}
 for a real power $b$. In this case, relation~(\ref{null:delta}) is
equivalent to
\begin{equation}
	\Delta f^\prime(R) = 0 ,
\label{null:deltaR}
\end{equation}
 at all orders of approximation for a power-law function of the Ricci
scalar
\begin{equation}
 f(R) = R^b.
\label{eq3b}
\end{equation}
 Substitution of function~(\ref{power-law}) into the null
variations of the gravitational field's action~(\ref{action}) in vacuum leads
to
\begin{equation}
  \delta S_\mathrm{f}  = - \frac{ c^3 }{ 16 \pi G } L_M^{2(b-1)} \delta\int{
     R^b \sqrt{-g} \, \mathrm{d}^4x} = 0 \, , 
\label{eq03}
\end{equation}
 and so
\begin{equation}
  \delta\int{ R^b \sqrt{-g} \, \mathrm{d}^4x } = 0 \, .
\label{eq03a}
\end{equation}
 From the last relation, we can see that the same field equations in vacuum are obtained for a power-law function~\eqref{power-law} in the $f(\chi)$ theory, as well as for a standard power-law $f(R)$ metric theory~\eqref{eq3b}, but with the important
restriction~\eqref{null:deltaR} needed to yield the correct relativistic
extension of MOND ($\chi \ll 1$ limit). \citet{mendoza13} showed that this condition is crucial to
describe the details observed for gravitational lensing for individuals,
groups and clusters of galaxies, and differs from the results obtained in
\citet{capozziello-newton}, for a standard $f(R)$ power-law description in
vacuum.  As discussed in \citet{mendoza13}, such discrepancy
occurs from the sign convention used in the definition of the Riemann tensor,
giving two different choices of signature that effectively bifurcate on the
solution space, a property which does not appear in Einstein's general
relativity.  This is due to higher order derivatives with respect to the
metric tensor that appear on metric theories of gravity (cf.~equations~\eqref{field:eqs} and~\eqref{trace}).  Following the results in
\citet{mendoza13}, we use the same definition of Riemann's tensor sign and the branch of solutions that
recover the correct weak field limit of the theory, in order to explain the rotation
curves of spiral galaxies based on the Tully-Fisher relation, and the
gravitational lensing observed at the
outer regions of groups and clusters of galaxies, within the point-mass description.

  Given the equivalence of the power-law $f(\chi)$ models with the standard $f(R)$ metric theories, the standard perturbation analysis for $f(R)$ theories constrained by equation~(\ref{null:deltaR}) is developed for the power-law
description of gravity~(\ref{power-law}) in the weak field limit, and for the first-order MOND-like
relativistic correction
in~\citet{mendoza13}. In this case, the standard field equations~\eqref{field:eqs} reduce to
\citep[see e.g.][]{capozziello-book}
\begin{equation}
	f^{\prime}(R) R_{\mu\nu} - \frac{1}{2} f(R) g_{\mu\nu} +
   \mathcal{H}_{\mu\nu} = 0 \, ,
\label{einstein:eqs}
\end{equation}
 where the fourth-order terms are grouped into
$\mathcal{H}_{\mu\nu} := - \left( \nabla_\mu \nabla_\nu
- g_{\mu\nu} \Delta \right) f^{\prime}(R)$. The trace of
equation~(\ref{einstein:eqs}) is given by
\begin{equation}
	f^{\prime}(R) R - 2f(R) + \mathcal{H} = 0 \, ,
\label{trace:einstein}
\end{equation}
 with $\mathcal{H} :=  \mathcal{H}_{\mu\nu}g^{\mu\nu} = 3 \Delta f^\prime(R)$.

  For the case of the static spherically symmetric space-time~(\ref{metric}),
it follows that
\begin{eqnarray}
  \mathcal{H}_{\mu\nu} &=& - f^{\prime\prime} \Big\{ R_{, \mu\nu} -
    \Gamma^1_{\mu\nu} R_{,r} \nonumber \\
   && \left. - g_{\mu\nu} \left[ \left(
    g^{11}_{~,r} + g^{11} \left( \ln{ \sqrt{-g} }
    \right)_{,r} \right) R_{,r} + g^{11} R_{,rr} \right] \right\} \nonumber \\
   && - f^{\prime\prime\prime} \left( R_{,\mu}R_{,\nu} -
    g_{\mu\nu}g^{11} R_{,r}^{~2} \right) \, ;
\label{H:exp}
\end{eqnarray}
and the trace
\begin{eqnarray}
  \mathcal{H} &=& 3 f^{\prime\prime} \left[ \left( g^{11}_{~,r} +
	g^{11} \left( \ln{\sqrt{-g}} \right)_{,r} \right) R_{,r}
	+ g^{11} R_{,rr} \right] \nonumber \\
    && + 3 f^{\prime\prime\prime} g^{11} R_{,r}^{~2} \, . 
\label{curlH:exp}
\end{eqnarray}

\section{Perturbation theory}

   In this subsection, we present the perturbation analysis for $f(\chi)$ metric theories. Perturbations applied to metric theories of gravity, including GR, are extensively detailed in the monograph by~\citet{will93}.  In
particular, for $f(R)$ metric theories, \citet{stabile09} have developed a second order
perturbation analysis and applied it to lenses and clusters of galaxies
\citep{capozziello-clusters}.

  The general field equations~(\ref{einstein:eqs})-(\ref{trace:einstein}) are of fourth order in the derivatives of the
metric tensor $g_{\mu\nu}$.  In dealing with the algebraic manipulations of the
perturbations of an $f(R)$ metric theory of gravity, T.~Bernal, S.~Mendoza and
L.A.~Torres developed a code in the Computer Algebra System (CAS) Maxima,
the MEXICAS (Metric Extended-gravity Incorporated through a Computer Algebraic
System) code (licensed with a GNU Public License Version 3).  The code is
described in \citet{mendoza13} and can be downloaded from:
\textit{http://www.mendozza.org/sergio/mexicas}. We use it to obtain the field
equations up to the fourth order in perturbations as described in the next subsections.

\subsection{Weakest field limit $\mathcal{O}(2)$ correction}
\label{lowest-order}

  Ricci's scalar can be written as follows:
\begin{equation}
	R = R^{(2)} + R^{(4)} + \mathcal{O}(6) \, ,
\label{ricci:exp}
\end{equation}
 which has non-null second and fourth perturbation orders in $v/c$ powers. The fact that
$ R^{(0)}=0 $ is consistent with the flatness of space-time assumption at
the lowest zeroth perturbation order. The $\mathcal{O}(2)$ term of Ricci's
scalar, $R^{(2)}$, from the metric components~\eqref{metric-perturbed:g00}-\eqref{metric-perturbed}, is given by
\begin{equation}
	R^{(2)} = - \frac{2}{r} \left[ g_{11,r}^{(2)} +
	\frac{g_{11}^{(2)}}{r} \right] - g_{00,rr}^{(2)} - \frac{2}{r}
	g_{00,r}^{(2)} \, .
\label{ricci:second}
\end{equation}

  At the lowest perturbation order, $\mathcal{O}(2b - 2)$, the 
trace~(\ref{trace:einstein}) for a power-law theory~(\ref{eq3b}) can be
written as \citep{mendoza13}
\begin{equation}
  \mathcal{H}^{(2b - 2)} = 3 \Delta f^{\prime (2b-2)} (R) = 0 .
\end{equation}
 Note that this is the only independent equation at this perturbation
order.  Substitution of~\eqref{metric:exp-upg00}, \eqref{metric:exp-up}, \eqref{eq3b} and
\eqref{ricci:exp} into the previous relation leads to a differential equation
for Ricci's scalar at order $ \mathcal{O}(2) $, which has the solution 
\citep{mendoza13}
\begin{equation}
  R^{(2)}(r) = \left[ (b-1) \left(\frac{\mathcal{A}}{r} + \mathcal{B} \right) 
    \right]^{1/(b-1)},
\label{R2:exact}
\end{equation}
 where $\mathcal{A}$ and $\mathcal{B}$ are constants of integration.
Far away from the central mass, space-time is flat and so, Ricci's scalar must
vanish at large distances from the origin, i.e.~the constant $\mathcal{B}=0$.

  Now, the case $b = 3/2$ has been shown to yield a MOND-like behavior in the limit $r \gg l_M \gg r_\mathrm{g}$ \citep{bernal11a,mendoza13}
and so, after substituting $b = 3/2$ and $\mathcal{B}=0$, solution~\eqref{R2:exact} becomes
\begin{equation}
\label{R2}
	R^{(2)}(r)  = \frac{\hat{R}}{r^2},
\end{equation}
 where the constant $\hat{R}:= \mathcal{A}^2/4$.

  At the next perturbation order, $\mathcal{O}(2b)$, the metric components
$ g_{00}^{(2)} $, $ g_{11}^{(2)} $, $ g_{00}^{(4)} $ and Ricci's scalar
$ R^{(4)}$ can be obtained. In this case, the field equations~\eqref{einstein:eqs} are given by \citep{mendoza13}
\begin{equation}
  b R^{(2) b -1} R_{\mu\nu}^{(2)} - \frac{1}{2}  R^{(2) b}
g_{\mu\nu}^{(0)} + \mathcal{H}^{(2b)}_{\mu\nu} = 0 \, ,
\label{eq7a:notes}
\end{equation}
where $\mathcal{H}^{(2b)}_{\mu\nu} = - \left( \nabla_\mu \nabla_\nu
- g_{\mu\nu} \Delta \right) f^{\prime (2b)}(R)$.
 From constraint~\eqref{null:deltaR} it follows that $ \Delta
f^{\prime (2b)} = 0 $, and last equation simplifies greatly. Using
relations~\eqref{metric-perturbed:g00}-\eqref{metric:exp-up} into the $00-$component of equation~\eqref{eq7a:notes} leads to \citep{mendoza13}
\begin{equation}
  b R^{(2)b-1} R_{00}^{(2)} - \frac{1}{2} R^{(2)b}  +  \frac{1}{2}
    b (b-1) g^{(2)}_{00,r} R^{(2)b-2} R^{(2)}_{,r} = 0 \, .
\label{eq11a:notes}
\end{equation}
The $00-$component of Ricci's tensor at $\mathcal{O}(2)$ is given by
\begin{equation}
  R^{(2)}_{00} = - \frac{r g^{(2)}_{00,rr} + 2 g^{(2)}_{00,r}} {2r} \, ;
\label{ricci:00}
\end{equation} 
 by substituting this last expression, $b = 3/2$, and
result~\eqref{R2} into equation~\eqref{eq11a:notes}, the following
differential equation for $g^{(2)}_{00}$ is obtained:
\begin{equation}
   r^2 g^{(2)}_{00,rr} + 3 r g^{(2)}_{00,r} + \frac{2 \hat{R}}{3} = 0 ,
\label{g00:diffeq}
\end{equation}
 which has the solution \citep{mendoza13}
\begin{equation}
    g^{(2)}_{00}(r) = - \frac{\hat{R}}{3}  \ln \left( \frac{r}{r_s} \right)
	+ \frac{ k_1 }{ r^2 } , 
\label{g00:notes}
\end{equation}
 where $k_1$ and $r_s$ are constants of integration. By
substitution of this last result and relation~\eqref{R2} into equation~\eqref{ricci:second}, the following differential equation for $g^{(2)}_{11}$
is obtained:
\begin{equation}
   r g^{(2)}_{11,r} + g_{11}^{(2)} + \frac{k_1}{r^2} + \frac{\hat{R}}{3} = 0,
\label{g11:diffeq}
\end{equation}
 with the following analytic solution \citep{mendoza13}:
\begin{equation}
  g^{(2)}_{11}(r) = \frac{k_1}{r^2} + \frac{k_2}{r} - \frac{\hat{R}}{3} ,
\label{grr2}
\end{equation} 
 where $k_2$ is another constant of integration.

   To fix the free parameters $\hat{R}$, $k_1$, $k_2$ in relations~\eqref{g00:notes}
and~\eqref{grr2}, \citet{mendoza13} compared the metric coefficients with
observations of rotation curves of spiral galaxies and the Tully-Fisher
relation, and with gravitational lensing results of individual, groups and
clusters of galaxies. They obtained:
\begin{eqnarray}
	g_{00}^{(2)} (r) &=& - \frac{2 \left( GMa_0\right)^{1/2}}{c^2} \ln \left( \frac{r}{r_s} \right) , \label{g00-mond} \\
    g_{11}^{(2)} (r) &=& - \frac{2 \left( GMa_0\right)^{1/2}}{c^2} ,
\end{eqnarray}
for $\hat{R}=6(GMa_0)^{1/2}/c^2$ and $k_1 = 0 = k_2$.
Their results are summarized in Table~\ref{table01}.
Notice that the metric component $g_{00}^{(2)}=2 \phi/c^2$ reduces to the MONDian gravitational potential, $\phi_\mathrm{MOND}=-(GMa_0)^{1/2} \ln(r/r_s)$, and to obtain the acceleration the length $r_s$ is left indeterminate at $\mathcal{O}(2)$.
However, as explained in Section~\ref{cluster-profiles}, its value is
necessary to describe the dynamics up to
$\mathcal{O}(4)$ of the theory, and it will be fixed with observational data.

Now, it is worth to notice the minus sign in $g_{00}^{(2)}$ (equation~\eqref{g00-mond}). To obtain the corresponding ``deep-MOND'' acceleration, $a_\mathrm{MOND}(r)=-(GMa_0)^{1/2}/r$, from such potential, we cannot use the standard definition $\bm{a}(r)=-\nabla\phi(r)$, as in Newtonian mechanics, since a positive gravitational potential can not produce bounded orbits in this theory \citep[cf.][]{Landau1982mechanics}. That is, we cannot choose $\phi_\mathrm{MOND}(r)=+(GMa_0)^{1/2} \ln (r/r_s)$, unlike the treatment followed in \citet{Campigotto:2017}. This fact explains their disagreement with the correct $(-)$ sign found in \citet{bernal11a,mendoza13}, and in this article.

\begin{table*}
\small
\centering
\caption{Empirical $\mathcal{O}(2)$ metric coefficients}
\begin{tabular}{|c|c|c|}
\hline       &   & 					\\
Metric coefficient & $g_{00}^{(2)}$ & $g_{11}^{(2)}$ \\
       &   & 					\\
\hline	&   & 					\\
& $- \frac{2 (GMa_0)^{1/2}}{c^2} \ln \left( \frac{r}{r_s} \right) $ &
$- \frac{2(GMa_0)^{1/2}}{c^2} $  		\\
Observations &   & 					\\
             &   (Tully-Fisher)      &  (Lensing)       \\
             &   & 					\\
\hline       &   & 					\\
Theory       & $-\frac{\hat{R}}{3}\ln \left( \frac{r}{r_s} \right) +
\frac{k_1}{r^2}$
& $ \frac{k_1}{r^2} + \frac{k_2}{r} - \frac{\hat{R}}{3} $ \\
$ f(\chi) = \chi^{3/2} $   &       &			\\
     &  $ \ \hat{R} = \frac{6(GMa_0)^{1/2}}{c^2} , $ \   $ k_1 = 0 \ $     &
$ \ \hat{R} = \frac{6(GMa_0)^{1/2}}{c^2} , \ k_1 = 0 = k_2 \ $ \\
&            &			\\
\hline
\end{tabular}
\begin{center}
The Table shows the results
obtained for the metric components $g_{00}^{(2)}$ and $g_{11}^{(2)}$
for a static spherically symmetric space-time. The metric coefficients are
empirically obtained from astronomical observations in the scales of galaxies
(Tully-Fisher relation) and lensing at the outer regions of individual, groups
and clusters of galaxies, and compared to the ones predicted by the $f(\chi) = \chi^{3/2}$ metric theory of gravity. Any proposed metric of a theory
of modified gravity must converge to the observational values presented in this
Table.  As shown in \citet{mendoza13}, the theory $ f(\chi) = \chi^{3/2} $ is in
perfect agreement with the metric components derived from observations.
\end{center}
\label{table01}
\end{table*}

\subsection{``Post-MONDian'' $\mathcal{O}(4)$ correction}

  In this subsection, we derive the first relativistic correction of the metric
theory $f(\chi)=\chi^{3/2}$, i.e.~we obtain $g_{00}^{(4)}$ and
$R^{(4)}$ having in mind further applications for material particles moving at high velocities compared to the speed of light.
Here, we assume the solutions for the $\mathcal{O}(2)$ metric coefficients as shown before and fit the $\mathcal{O}(4)$ ones as explained below.

 The $\mathcal{O}(4)$ component of Ricci's scalar is given by
\begin{eqnarray}
   R^{(4)} &=& - \frac{2}{r} \left\{ g_{11}^{(2)} \left[ 2 g_{11,r}^{(2)} +
   g_{00,r}^{(2)} + \frac{g_{11}^{(2)}}{r} \right]  \right. \nonumber \\
   && + g_{00,r}^{(2)} \left[
   \frac{r}{4} \left( g_{11,r}^{(2)} - g_{00,r}^{(2)} \right) - g_{00}^{(2)} \right] + g_{00,r}^{(4)} \Bigg\} \nonumber \\
   && + g_{00,rr}^{(2)} \left[ g_{00}^{(2)} - g_{11}^{(2)} \right]
    - g_{00,rr}^{(4)} \, . 
\label{R4:metric}
\end{eqnarray}
  To obtain the $\mathcal{O}(4)$ metric coefficients, we use another independent field equation. Substitution of 
relations~\eqref{metric-perturbed:g00}-\eqref{metric:exp-up} into the $22-$component of equation~\eqref{eq7a:notes} yields
\begin{eqnarray}
  & b (b-1) r R^{(2)b-2} \left[ R^{(4)}_{,r} + g_{11}^{(2)}
  R^{(2)}_{,r} + (b-2) R^{(2)-1} R^{(2)}_{,r} R^{(4)} \right] \nonumber \\
  & - b R^{(2)b-1} R_{22}^{(2)} - \frac{r^2}{2} R^{(2)b} = 0 ,
\label{G:22}
\end{eqnarray}
 where the $22-$component of the Ricci tensor at order
$\mathcal{O}(2)$ is given by
\begin{equation}
  R^{(2)}_{22} = g_{11}^{(2)} + \frac{r}{2} \left[ g^{(2)}_{00,r} +
  g^{(2)}_{11,r} \right] .
\label{ricci:22}
\end{equation} 
 By substitution of the last
equation together with solutions~\eqref{R2}, \eqref{g00:notes} and~\eqref{grr2} for $R^{(2)}$, $g_{00}^{(2)}$ and $g_{11}^{(2)}$, respectively, into
equation~\eqref{G:22} for $b=3/2$, we obtain the following differential equation for 
Ricci's scalar at $\mathcal{O}(4)$:
\begin{equation}
   r^4 R^{(4)}_{,r} + r^3 R^{(4)} + \hat{R}^2 r - 3 k_2 \hat{R} = 0 ,
\label{dR4}
\end{equation}
 which has the following exact solution:
\begin{equation}
   R^{(4)}(r) = \frac{\hat{R}^2}{r^2} - \frac{3 k_2 \hat{R}}{2 r^3} -
   \frac{4 k_3}{c^4 r} ,
\label{R4}
\end{equation}
 where $k_3$ is a constant of integration.
 
 Now, from the definition~\eqref{R4:metric} of Ricci's scalar $R^{(4)}$ 
and from equation~\eqref{R4}, together with~\eqref{R2}, \eqref{g00:notes}
and~\eqref{grr2}, we obtain the following differential equation for
$g^{(4)}_{00}$:
\begin{eqnarray}
   &&- 9 g_{00,rr}^{(4)} - \frac{18}{r} \left( g_{00,r}^{(4)} - \frac{2 k_3}{c^4} \right)
   + \frac{\hat{R}^2}{r^2} \left[ \ln \left( \frac{r}{r_s} \right) -
   \frac{23}{2} \right] \nonumber \\
   && + \frac{3 k_2}{r^3} \left( 5 \hat{R} +
   \frac{6 k_2}{r} \right)
   - \frac{3 k_1 \hat{R}}{r^4} \left[ 2 \ln \left( \frac{r}{r_s} \right) +
   1 \right] \nonumber \\
   && + \frac{45 k_1 k_2}{r^5} + \frac{54 k_1^2}{r^6} = 0 \, ,
\label{g004:diffeq}
\end{eqnarray}
 with the exact solution for $g^{(4)}_{00}$:
\begin{eqnarray}
   g^{(4)}_{00}(r) &=& \frac{\hat{R}^2}{18} \ln^2 \left( \frac{r}{r_s}\right) -
   \frac{25 \hat{R}^2}{18} \ln \left( \frac{r}{r_s}\right) + \frac{2 k_3}{c^4}r  \nonumber \\
   && + \frac{2 k_4}{c^4 r} + \frac{k_5}{c^4} - \frac{k_1 \,\hat{R}}{3\,{r}^{2}} \left[ \ln\left(\frac{r}{r_s}\right) +
   2 \right] + \frac{{k_2}^{2}}{{r}^{2}} \nonumber \\
   && - \frac{5\,k_2\,\hat{R}}{3\,r} \left[ \ln\left(\frac{r}{r_s}\right) + 1
   \right] 
   + \frac{5\,k_1\,k_2}{6\,{r}^{3}} +
   \frac{{k_1}^{2}}{2\,{r}^{4}} \, , \nonumber \\
\label{g004}
\end{eqnarray}
 where $k_4$ and $k_5$ are constants of integration.

  Now, by using the same results for the parameters $\hat{R}=6(GMa_0)^{1/2}/c^2$, $k_1=0=k_2$, from \citet{mendoza13} (see Table~\ref{table01}),
the metric coefficient $g_{00}^{(4)}$ and Ricci's scalar $R^{(4)}$ reduce to
\begin{eqnarray}
   g_{00}^{(4)} &=& \frac{2GMa_0}{c^4} \ln \left(
        \frac{r}{r_s}\right) \left[\ln \left(\frac{r}{r_s}\right) - 25
        \right] \nonumber \\
   && + \frac{2 k_3}{c^4} r + \frac{2 k_4}{c^4 r} + \frac{k_5}{c^4} \, ;\\
\label{g004-R4}
   R^{(4)} &=& \frac{36 GMa_0}{c^4} \frac{1}{r^2} -
        \frac{4 k_3}{c^4 r} \, .
\end{eqnarray}
To fix the constants $k_3$ and $k_4$ ($k_5$ vanishes upon
derivation of $g_{00}^{(4)}$ with respect to $r$ (cf.~equation~\eqref{v:fourth}), we fit the observational data of 12 clusters
of galaxies, as described in Section~\ref{cluster-profiles}.

\section{Generalization to extended systems}
\label{gen-systems}

In \citet{mendoza13}, the details of gravitational lensing for individual, groups and clusters of galaxies at the outer regions of those systems were obtained considering a point-mass lens. \citet{Campigotto:2017} compared those results with specific observations of gravitational lensing and found a large discrepancy between the $\mathcal{O}(2)$ terms in the metric coefficients of the $f(\chi)=\chi^{3/2}$ gravity and the observations. However, to obtain the correct gravitational lensing it is necessary to take into account the mass-density distribution, using a suitable gravitational potential.

In this section, we assume the solution for the MONDian point-mass gravitational potential for the $f(\chi)=\chi^{3/2}$ model, and generalize it to spherically symmetric mass distributions through potential theory. To this aim, we take into account the potential due to differentials of mass and integrate for the interior and exterior shells for a given radius $r$.

  Let us take the radial component~(\ref{radial:geodesic}) of geodesic equations~(\ref{geodesic}) in the weak field limit of the theory.  In
this limit, the rotation curve for test particles bounded to a circular orbit
about a mass $M$ with circular velocity $v(r)$ is given by
equation~(\ref{centrifugal}). Such equation, up to $\mathcal{O}(4)$ of
approximation, is given by
\begin{equation}
    \frac{a_c}{c^2} = \frac{1}{c^2} \frac{\mathrm{d}^2 r}{\mathrm{d} t^2} = 
     \frac{1}{2} \left[
	 g^{(2)}_{00,r} + g^{(2)}_{11} g^{(2)}_{00,r} + g^{(4)}_{00,r}
	 \right].
\label{v:fourth}
\end{equation}

  Substitution of the $\mathcal{O}(2)$ results for the metric
coefficients, $g_{00}^{(2)}$ and $g_{11}^{(2)}$ (see Table~\ref{table01}), and solution~\eqref{g004} for $g_{00}^{(4)}$
in equation~\eqref{v:fourth}, results in the following expression for the acceleration
of a test-mass particle in the gravitational field generated by the point-mass
$M$:
\begin{eqnarray}
   && a_c(r) = - \frac{\left(G M a_0\right)^{1/2}}{r} \label{acc:fourth} \\
    && - \frac{1}{c^2} \left[ \frac{23 G M a_0}{r} - \frac{2 G M a_0}{r}
	\ln \left( \frac{r}{r_s} \right) - k_3 +
	\frac{k_4}{r^2} \right] . \nonumber
\end{eqnarray}
The first term on the right-hand
side of last equation corresponds to the ``deep-MOND'' acceleration. The
remaining $\mathcal{O}(2)$ terms are the
first relativistic correction to the gravitational acceleration.

   In order to apply these results to extended systems, it is necessary to
generalize the gravitational acceleration~\eqref{acc:fourth} to a spherical
mass distribution $M(r)$. To do this, notice that the first term of such equation can be easily
generalized: In this case, the deep-MONDian
acceleration can be written as $f(x)= a/a_0 =x$, for $x:=l_M/r$. As discussed
in \citet{mendoza11}, this function, and in general any analytic function
which depends only on the parameter $x$, guarantees Newton's theorems.  In
other words, the gravitational acceleration exerted by the outer shells at
position $r$ cancels out and depends only on the mass $M(r)$ interior to $r$.
Thus, the first MONDian term of the gravitational acceleration~\eqref{acc:fourth} due to a mass
distribution can be written as
\begin{equation}
   a_c(r) = - \frac{\left[ G M(r) a_0 \right]^{1/2}}{r}.
\label{deepMOND-acc}
\end{equation}

   For the $\mathcal{O}(2)$ terms on the right-hand side of
acceleration~\eqref{acc:fourth}, let us take the corresponding gravitational potential generated by the point-mass $M$.
After integrating the $\mathcal{O}(2)$ terms with respect to the radius
$r$, according to the spherical symmetry assumption, we obtain
\begin{eqnarray}
   \phi^{(2)} (r) &=& - \frac{G M a_0}{c^2} \left\{  - \frac{A}{r} - Br  \right. \label{grav-potential2} \\
   && +\ln \left( \frac{r}{r_s}
	\right) \left[ 23 - \ln \left( \frac{r}{r_s} \right) \right] \bigg\} \, , \nonumber 
\end{eqnarray}
 where we have assumed $k_3$ and $k_4$ are proportional to $GMa_0$ and we have defined, for convenience, the constants $A:=k_4/GMa_0$ and $B:=k_3/GMa_0$.
 This point-mass gravitational potential can be generalized
considering that the extended system is composed of many infinitesimal mass
elements $\mathrm{d}M$, each one contributing with a point-like gravitational
potential~\eqref{grav-potential2}, such that
\begin{equation}
	M(r) = \int_V \mathrm{d}M = \int_V \rho(r') \, \mathrm{d}V' ,
\label{mass-rho}
\end{equation}
where $\rho$ is the mass-density of the system and the volume element is $\mathrm{d}V'=r'^2 \, \sin{\theta'} \,
\mathrm{d}\varphi' \, \mathrm{d}\theta' \, \mathrm{d}r'$, integrated over the
volume $V$. 

  From equation~\eqref{grav-potential2}, the generalized gravitational
potential in spherical symmetry is the convolution
\begin{equation}
	\int{ f(\mathbf{r}-\mathbf{r'}) \rho(r') r'^2 \sin{\theta'}
	\mathrm{d}\varphi' \mathrm{d}\theta' \mathrm{d}r' },
\end{equation}
 of the function
\begin{eqnarray}
   f(\mathbf{r}-\mathbf{r'}) &=& - \frac{G a_0}{c^2}
	\left\{  - \frac{A}{|\mathbf{r}-
	\mathbf{r'}|} - B|\mathbf{r}-\mathbf{r'}| \right. \\
    && \left. + \ln \left( \frac{|\mathbf{r}-\mathbf{r'}|}{r_s} \right) \left[
	23 - \ln \left( \frac{|\mathbf{r}-\mathbf{r'}|}{r_s} \right) \right] \right\} , \nonumber 
\end{eqnarray}
 with the differential $\mathrm{d}M$ defined in~(\ref{mass-rho}), for $f$ and
$\rho$ locally integrable functions for $r > 0$ \citep[see e.g.][]{vladimirov2002}.
Due to the spherical symmetry of the problem, the integration can be done in one direction, for
example the $z$ axis, where the polar angle $ \theta = 0$ and
$|\mathbf{r}-\mathbf{r'}|=\sqrt{r^2+r'^2-2rr'\cos\theta'}$.  Thus, the
$\mathcal{O}(2)$ generalized gravitational potential for a mass distribution
can be written as
\begin{eqnarray}
   \Phi^{(2)}(r) &=& - \frac{G a_0}{c^2} \int_0^\mathcal{R} \int_0^\pi
	\int_0^{2 \pi} \left\{ - \frac{A}{|\mathbf{r}-
	\mathbf{r'}|} \right. \\
	&& + \ln \frac{|\mathbf{r}-\mathbf{r'}|}{r_s}
	\left[ 23 - \ln \frac{|\mathbf{r}-\mathbf{r'}|}{r_s} \right] \nonumber \\
	&& - B |\mathbf{r}-\mathbf{r'}| \bigg\} \rho(r') \, r'^2 \, \sin{\theta'} \,
	\mathrm{d}\varphi' \, \mathrm{d}\theta' \, \mathrm{d}r' , \nonumber
\label{grav-potential3}
\end{eqnarray}
 integrated over the whole volume $V$. If the density distribution is
known, the generalized potential~(\ref{grav-potential3}) can be numerically
integrated to obtain the gravitational acceleration, from $0<r<r'$ and
$r'<r<\mathcal{R}$, where $\mathcal{R}$ is the radius of the spherical
configuration.

  Notice that the term with constant $A$ on the last integral is a Newtonian-like potential, and it is a well-known result that the matter outside the
spherical shell of radius $r$ does not contribute to the corresponding gravitational acceleration, thus we have
\begin{equation}
  a_c^{(2)}(r) = - \frac{G M(r) a_0 A}{c^2 r^2} .
\end{equation}
For the other terms, the integration is
done for the interior and exterior shells of mass $\mathrm{d}M$ with respect to the radius
$r$, giving as result the following expression:
\begin{eqnarray}
   &\Phi_{c}^{(2)}(r) = - \frac{2 \pi G a_0}{c^2 r} \int_0^\mathcal{R} \left\{ - \frac{B}{3} \left[ \left(r+r'\right)^3 - \left|r-r'\right|^3
	\right]
	\right. \nonumber \\
    & \left. + (r-r')^2 \ln \left( \frac{|r-r'|}{r_s} \right) \left[ \frac{1}{2} \ln \left(
	\frac{|r-r'|}{r_s} \right) - 24 \right] \right. \nonumber \\
	& \left. - (r+r')^2 \ln \left( \frac{r+r'}{r_s} \right) \left[ \frac{1}{2} \ln \left(
	\frac{r+r'}{r_s} \right) - 24 \right] 
	 \right\} \rho(r') r' \, \mathrm{d}r' \nonumber \\
	& + \frac{12 G a_0}{c^2} \int_0^\mathcal{R} 4 \pi \rho(r') r'^2 \,
	\mathrm{d}r' ,
\label{Phi_c1}
\end{eqnarray}
where the last term is constant. Now, after performing the derivation of the
potential~(\ref{grav-potential3}) with respect to $r$ and simplifying some
terms, the generalized gravitational acceleration for a spherical mass distribution $M(r)$ can be written as 
\begin{eqnarray}
   a_c(r) &=& - \frac{\left[ G M(r) a_0 \right]^{1/2}}{r} +
	\frac{\mathrm{d}\Phi^{(2)}(r)}{\mathrm{d}r}, \label{final-acc} \\
	  &=& - \frac{\left[ G M(r) a_0 \right]^{1/2}}{r} -
	\frac{G M(r) a_0 A}{c^2 r^2}
	+ \frac{\mathrm{d}\Phi_c^{(2)}(r)}{\mathrm{d}r} \, , \nonumber
\end{eqnarray}
 which can be obtained for an arbitrary density profile $\rho(r)$. Notice that the
parameters $B$ and $r_s$ appear only on the last term of
last equation through~(\ref{Phi_c1}).

\section{Fit with observations of clusters of galaxies}
\label{cluster-profiles}

  To compare the correction $g_{00}^{(4)}$ with the observations of clusters of
galaxies, we suppose the $\mathcal{O}(4)$ terms might be important in order to describe these systems, since their observed typical velocity dispersions are of the order $10^{-4}-10^{-3}$ times the speed of light. Also, as shown in \citet{sadeh-lahav2015,wojtak2011}, there exist observational relativistic effects of the velocity of the galaxies at the edge of the clusters, showing a difference of the inferred background potential with the galaxy inferred potential. Here we prove if the dynamical masses of the clusters can be explained with the extra $\mathcal{O}(4)$ terms from the metric coefficients of the $f(\chi)=\chi^{3/2}$ theory, without the necessity of DM.

\subsection{Galaxy clusters mass determination}
\label{fit-observations}

  To apply the results of the last subsection to the spherically symmetric
X-ray clusters of galaxies reported in~\citet{vikhlinin06}, notice that there
are two observables: the ionized gas profile $\rho_\mathrm{g}(r)$ and the
temperature profile $T(r)$. Under the hypothesis of hydrostatic equilibrium,
the hydrodynamic equation can be derived from the collisionless isotropic
Boltzmann equation for spherically symmetric systems in the weak field limit
\citep{binney-tremaine}:
\begin{equation}
   \frac{\mathrm{d}\left[ \sigma^2_r \rho_\mathrm{g}(r) \right]}{\mathrm{d}r}
   + \frac{\rho_\mathrm{g}(r)}{r} \left[ 2 \sigma_r^2 - \left(
   \sigma^2_{\theta} + \sigma_{\varphi}^2 \right) \right] =
   - \rho_\mathrm{g}(r) \frac{\mathrm{d} \Phi(r)}{\mathrm{d}r} ,
\label{boltzmann}
\end{equation}
 where $\Phi$ is the gravitational potential and $\sigma_r$,
$\sigma_{\theta}$ and $\sigma_{\varphi}$ are the mass-weighted velocity
dispersions in the radial and tangential directions, respectively. For an
isotropic system with rotational symmetry there is no preferred transverse
direction, and so $\sigma_{\theta} = \sigma_{\varphi}$.  For an isotropic
distribution of the velocities, we also have $\sigma_r = \sigma_{\theta}$.

 The radial velocity dispersion can be related to the pressure profile
$P(r)$, the gas mass density $\rho_{\mathrm{g}}(r)$ and the temperature
profile $T(r)$ by means of the ideal gas law to obtain:
\begin{equation}
   \sigma_r^2 = \frac{P(r)}{\rho_{\mathrm{g}}(r)} =
	\frac{k_{\mathrm{B}} T(r)}{\mu m_p},
\label{pressure-rhog}
\end{equation}
 where $k_{\mathrm{B}}$ is the Boltzmann constant,
$\mu = 0.5954$ is the mean molecular mass per particle for primordial He abundance and $m_p$ is the
proton mass.  Direct substitution of equation~(\ref{pressure-rhog})
into~(\ref{boltzmann}) yields the gravitational equilibrium relation:
\begin{equation}
   |\bm{a}(r)| = \left| \frac{\mathrm{d} \Phi(r)}{\mathrm{d}r}
   \right| = \frac{k_{\mathrm{B}} T(r)}{\mu m_p r}
   \left[ \frac{\mathrm{d} \ln \rho_\mathrm{g}(r)}{\mathrm{d}\ln r} +
   \frac{\mathrm{d} \ln T(r)}{\mathrm{d}\ln r} \right] .
\label{Beq-T-rho}
\end{equation}
  The right-hand side of the previous equation is determined by observational
data, while the left-hand side should be consistent through a given
gravitational acceleration and distribution of matter.

  In standard Newtonian
gravity, the total mass of a galaxy cluster is given by the mass of the gas, $M_\mathrm{gas}$, and the stellar
mass of the galaxies, $M_\mathrm{stars}$, inside it, with the necessary addition of an unknown DM
component to avoid a discrepancy of one order of magnitude on both sides of last equation.
In this case, the ``dynamical'' mass of the system,
$M_\mathrm{dyn}$, is determined by Newton's acceleration,
$a_\mathrm{N}$, as follows:
\begin{eqnarray}
\label{Newt-mass}
   M_\mathrm{dyn}(r) &:=& - \frac{r^2 a_\mathrm{N}(r)}{G}  \\
		    &=& - \frac{k_{\mathrm B} T(r)}{\mu m_p G} r \left[ \frac{
			\mathrm{d} \ln \rho_\mathrm{g}(r)}{\mathrm{d}\ln r} +
			\frac{\mathrm{d} \ln T(r)}{\mathrm{d}\ln r}
			\right]. \nonumber
\end{eqnarray}

 In the $ f(\chi)=\chi^{3/2} $ model, the acceleration will be
given by equation~\eqref{final-acc}. In this case, we define the ``theoretical''
mass, $M_{\mathrm{th}}$, as
\begin{eqnarray}
\label{M-teo}
   M_\mathrm{th}(r) &:=& - \frac{r^2 a_c(r)}{G} \nonumber \\
	&=& \left[ \frac{M_\mathrm{b}(r) a_0}{G} \right]^{1/2} r + \frac{M_\mathrm{b}(r) a_0 A}{c^2} \nonumber \\
	&& - \frac{r^2}{G} \frac{\mathrm{d}\Phi_c^{(2)}(r)}{\mathrm{d}r} ,
\end{eqnarray}
 where the baryonic mass of the system, $M_\text{b}(r)$, is given by
\begin{equation}
   M_\text{b}(r) = M_\mathrm{gas}(r) + M_\mathrm{stars}(r).
\label{M-bar}
\end{equation}

   In order to reproduce the observations, the theoretical mass obtained from
our modification to the gravitational acceleration must be equal to the
dynamical mass coming from observations, i.e.~$M_\mathrm{th} = M_\mathrm{dyn}$, without the inclusion of DM. This provides an observational
procedure to fit the three free parameters of our model ($r_s$, $A$ and $B$).

\subsection{\textit{Chandra} clusters sample}
\label{chandra-sample}

For this work, we used 12 X-ray galaxy clusters from the \textit{Chandra} Observatory, analyzed in \citet{Vikhlinin:2005,vikhlinin06}. It is a representative sample of low-redshift ($z \sim 0.01-0.2$, with median $z=0.06$), relaxed clusters, with very regular X-ray morphology and weak signs of dynamical activity. The effect of evolution is small within such redshift interval \citep{vikhlinin06}, thus we did not include the effects of the expansion of the Universe in our analysis. The observations extend to a large fraction of the virial radii, with total masses $M_{500}\footnote{$M_{500}$ is the mass at the radius $r_{500}$, where the density is $500 \rho_\mathrm{crit}$, with $\rho_\mathrm{crit}$ the critical density of the Universe.} \sim (0.5-10) \times 10^{14} \ \mathrm{M}_\odot$, thus the obtained values of the clusters properties (gas density, temperature and total mass profiles) are reliable \citep{vikhlinin06}.

   The keV temperatures observed in clusters of galaxies are interpreted in
such a way that the gas is fully ionized and the hot plasma is mainly emitted
by free-free radiation processes. There is also line emission by the ionized
heavy elements.  The radiation process generated by these mechanisms is
proportional to the emission measure profile $n_p n_e(r)$. \citet{vikhlinin06}
introduced a modification to the standard $\beta$-model
\citep{cavaliere1978}, in order to reproduce the observed features from the
surface brightness profiles, the gas density at the centers of relaxed
clusters and the observed X-ray brightness profiles at large radii.  A second
$\beta$-model component (with small core radius) is added to increase
accuracy near the clusters centers.  With these
modifications, the complete expression for the emission measure profile has 9
free parameters.  The 12 clusters can be adequately fitted by this model.  The
best fit values to the emission measure for the 12 clusters of galaxies can be
found in Table 2 of~\citet{vikhlinin06}.

   To obtain the baryonic density of the gas, the primordial abundance of He
and the relative metallicity $Z = 0.2 \ \mathrm{Z}_\odot$ are taken into account, and so
\begin{equation}
   \rho_\mathrm{g}(r) = 1.624 m_p \sqrt{n_p n_e (r)} .
\label{rho_gas}
\end{equation}

  In order to have an estimation of the stellar component of the clusters, we used the empirical relation
between the stellar and the total mass (baryonic + DM) in the
Newtonian approximation \citep{lin-vikhlinin12}:
\begin{equation}
   \frac{M_{\mathrm{stars}}}{10^{12} \mathrm{M}_\odot} = (1.8 \pm 0.1) \left(
	\frac{M_{500}} {10^{14} \mathrm{M}_\odot} \right)^{0.71 \pm 0.04}.
\label{stellar-mass}
\end{equation}
However, the total stellar mass is $\sim 1 \%$ of the total mass of the clusters, so we simply estimate the baryonic mass with the gas mass.

   For the temperature profile $T(r)$, \citet{vikhlinin06} used a different
approach from the polytropic law to model non-constant cluster temperature
profiles at large radii. All the projected temperature profiles show a broad
peak near the centers and decrease at larger radii, with a temperature
decline toward the cluster center, probably because of the presence of
radiative cooling \citep{vikhlinin06}. To model the temperature profile in
three dimensions, they constructed an analytic function such that outside the
central cooling region the temperature profile can be represented as a broken
power law with a transition region:
\begin{equation}
    T(r) = T_0 \frac{x+T_\text{min}/T_0}{1+x}\frac{(r/r_t)^{-a}}{\left[ 1 + (r/r_t)^b\right]^{c/b}} ,
\label{temp-profile}
\end{equation}
where $x:=(r/r_\text{cool})^{a_\text{cool}}$. The 8 best-fit parameters ($a,\,b,\,c,\,r_t,\,T_0,\,T_\text{min},\,r_\text{cool},\,a_\text{cool}$) for the
12 clusters of galaxies can be found in Table 3
of \citet{vikhlinin06}.

The total dynamical masses, obtained with equation~\eqref{Newt-mass} from the derived gas densities~\eqref{rho_gas} and temperature
profiles~\eqref{temp-profile} for the 12 galaxy clusters, were kindly
provided by Alexey Vikhlinin, along with the $1\sigma$ confidence levels from their Markov Chain Monte Carlo simulations. We used such data to fit our model as described in the next subsection.

\subsection{Parameters estimation method}

We conceptualized the free parameters calibration, $A$, $B$ and $r_s$, as
an optimization problem and proposed to solve it using Genetic Algorithms
(GAs), which are evolutionary based stochastic search algorithms that, in some
sense, mimic natural evolution.  In this heuristic technique, points in
the search space are considered as individuals (solution candidates), which as
a whole form a population.  The particular fitness of an individual is a
number indicating their quality for the problem at hand. As in nature, GAs
include a set of fundamental genetic operations that work on the genotype
(solution candidate codification): mutation, recombination and selection
operators \citep{Mitchell-98}.

  These algorithms operate with a population of individuals $P(t) =
{x_{1}^{t}, ...,x_{N}^{t}}$, for the $t$-th iteration, where the fitness of
each $x_{i}$ individual is evaluated according to a set of objective
functions $f_{j}(x_{i})$. These objective functions allow to order individuals of the population in a continuum of degrees of
adaptation.  Then, individuals with higher fitness recombine their
genotypes to form the gene pool of the next generation, in which random
mutations are also introduced to produce new variability.

 A fundamental advantage of GAs versus traditional methods is that GAs solve
discrete, non-convex, discontinuous, and non-smooth problems successfully, and
thus they have been widely used in many fields, including astrophysics and cosmology \citep[see e.g.][]{Charbonneau:1995,Canto:2009,Nesseris:2010,Curiel:2011,Rajpaul:2012,Lopez-Corona:2015}.

  It is important to note that, as it is well known from Taylor series, any
(normal) function may be well approximated by a polynomial, up to certain
correct order of approximation. Of course, even that this is correct from a
mathematical point of view, it is possible to consider that a polynomial
approximation is not universal for any physical phenomenon.  In this line of
thought, one may fit any data using a model with many free parameters, and
even in this approximation we may have a great performance in a statistical
sense, but it could be incorrect from the physical perspective.

  In this sense, an important question to ask is: How much better a complex
model is in a fitting process, justifying the
incorporation of extra parameters? In a more straightforward sense, how do we
carry out a fit with simplicity? Such question has been the motivation in the
recent years for new model selection criteria development in statistics, all
of them defining simplicity in terms of the number of parameters or the
dimension of a model \citep[see e.g.][for a non-technical introduction]{Forster-94}.  These criteria include Akaike's Information Criterion (AIC)
\citep{Akaike-74,Akaike-85}, the Bayesian Information Criterion (BIC)
\citep{Schwarz-78} and the Minimum Description Length (MDL)
\citep{Rissanen-89}. They fit the parameters of a model a little different
between them, but all of them address the same problem as a significance
test: Which of the estimated ``curves'' from competing models best represent
reality? \citep{Forster-94}.

 \citet{Akaike-74,Akaike-85} has shown that choosing the model with the
lowest expected information loss (i.e., the model that minimizes the
expected Kullback-Leibler discrepancy) is asymptotically equivalent to
choosing a model $M_{j}$, from a set of models $j=1,2,...,k$, that has the
lowest AIC value, defined by
\begin{equation}
 \mathrm{AIC}= - 2 \ln\left(\mathcal{L}_{j}\right)+2V_{j},
\end{equation}
where $\mathcal{L}_{j}$ is the maximum likelihood for the candidate
model and is determined by adjusting the $V_{j}$ free parameters in such a way
that they maximize the probability that the candidate model has generated the
observed data.  This equation shows that AIC rewards descriptive accuracy via
the maximum likelihood, and penalizes lack of parsimony according to the
number of free parameters (note that models with smaller AIC values are to be
preferred).  In that sense, \citet{Akaike-74,Akaike-85} extended this paradigm
by considering a framework in which the model dimension is also unknown, and
must therefore be determined from the data.  Thus, Akaike proposed a framework
where both model estimation and selection could be simultaneously
accomplished. For those reasons, AIC is generally regarded as the first, most
widely known and used model selection tool.

  Taking as objective function the AIC information index, we performed a
GAs analysis using a modified version of the \citet{Kumara} code
in C++.  The GA we used evaluates numerically equation~(\ref{M-teo}) in order
to compare the numerical results from the theoretical model with the cluster observational
data, as explained in Subsection~\ref{chandra-sample}.
All parameters were searched in a broad range from $ -1 \times 10^4 $  to
$ 1 \times 10^{10} $, generating populations of 1,000 possible solutions
over a maximum of 500,000 generation search processes.  We selected standard
GAs: tournament selection with replacement \citep{goldberg,
sastry}, simulated binary crossover \citep{deb1,deb2} and polynomial
mutation \citep{deb1,deb2,deb3}.  The parameters were estimated taking the
average from the first best population decile, checking the consistency of the $\mathcal{O}(2)$ corrections with respect to the zeroth order term in the gravitational
acceleration~(\ref{final-acc}).  Finally, since we obtained
\begin{equation}
	\Delta_{\mathrm{AIC}}:=\mathrm{AIC}_{i}-\mathrm{min}\left\{
	\mathrm{AIC}_{i}\right\} < 2 ,
\end{equation}
 for the parameters estimation,
then the model was accepted as a good one \citep{burnham}.

\section{Results and discussion}
\label{discussion}

\begin{figure*}
\caption{{\bf Dynamical mass vs. radius for the galaxy clusters.}}
  \includegraphics[width=\textwidth]{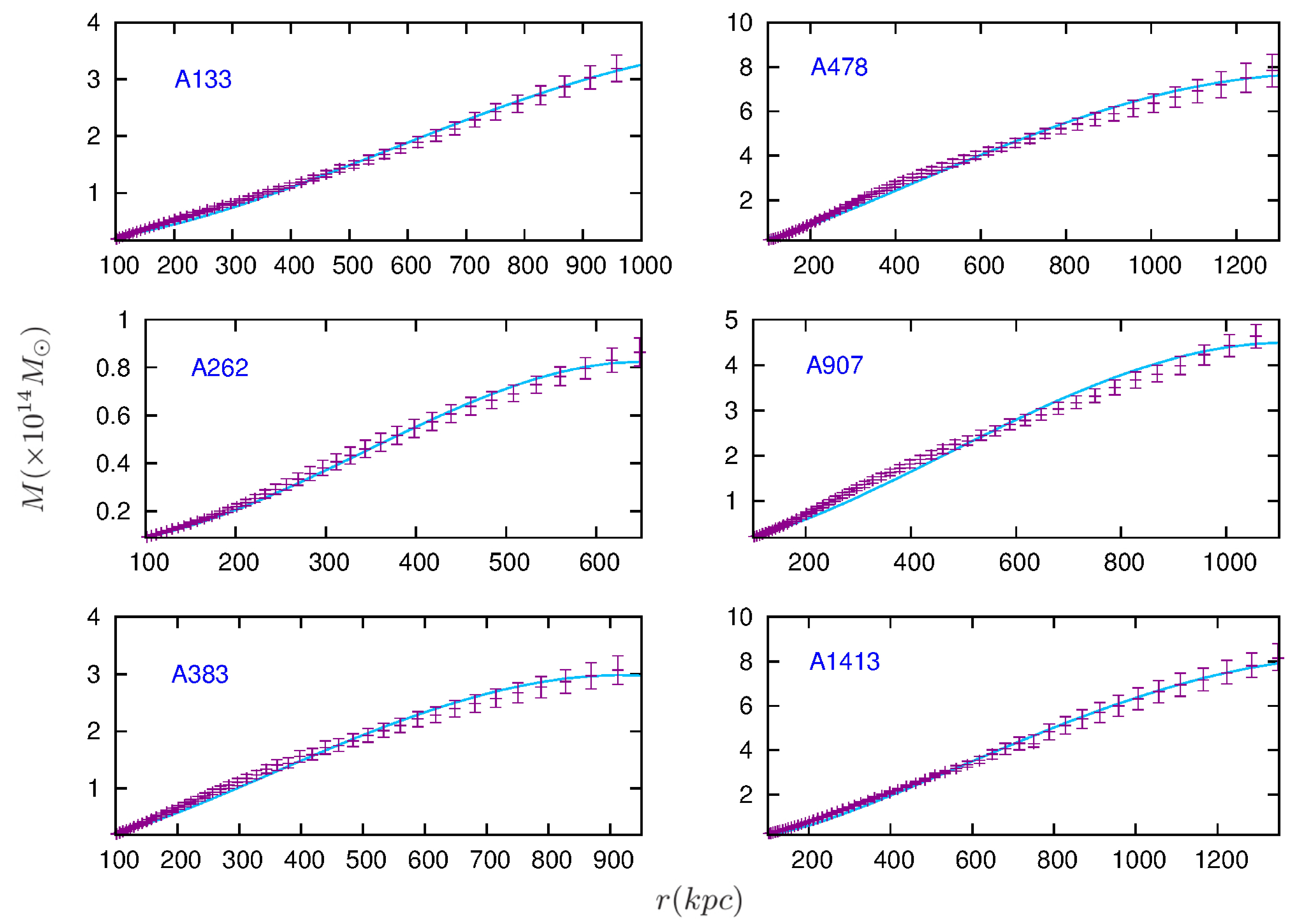}
\begin{center}  {Dynamical mass vs.~radius for the first 6 clusters of galaxies,
  with the best-fit parameters as summarized in Table~\ref{table01}. The
  points with uncertainty bars are the $1\sigma$ dynamical masses obtained in~\citet{vikhlinin06}. The solid line is the best fit obtained with our
  model.}
 \end{center}
\label{plot_data1-6}
\end{figure*}

\begin{figure*}
\caption{{\bf Dynamical mass vs.~radius for the galaxy clusters.}}
  \includegraphics[width=\textwidth]{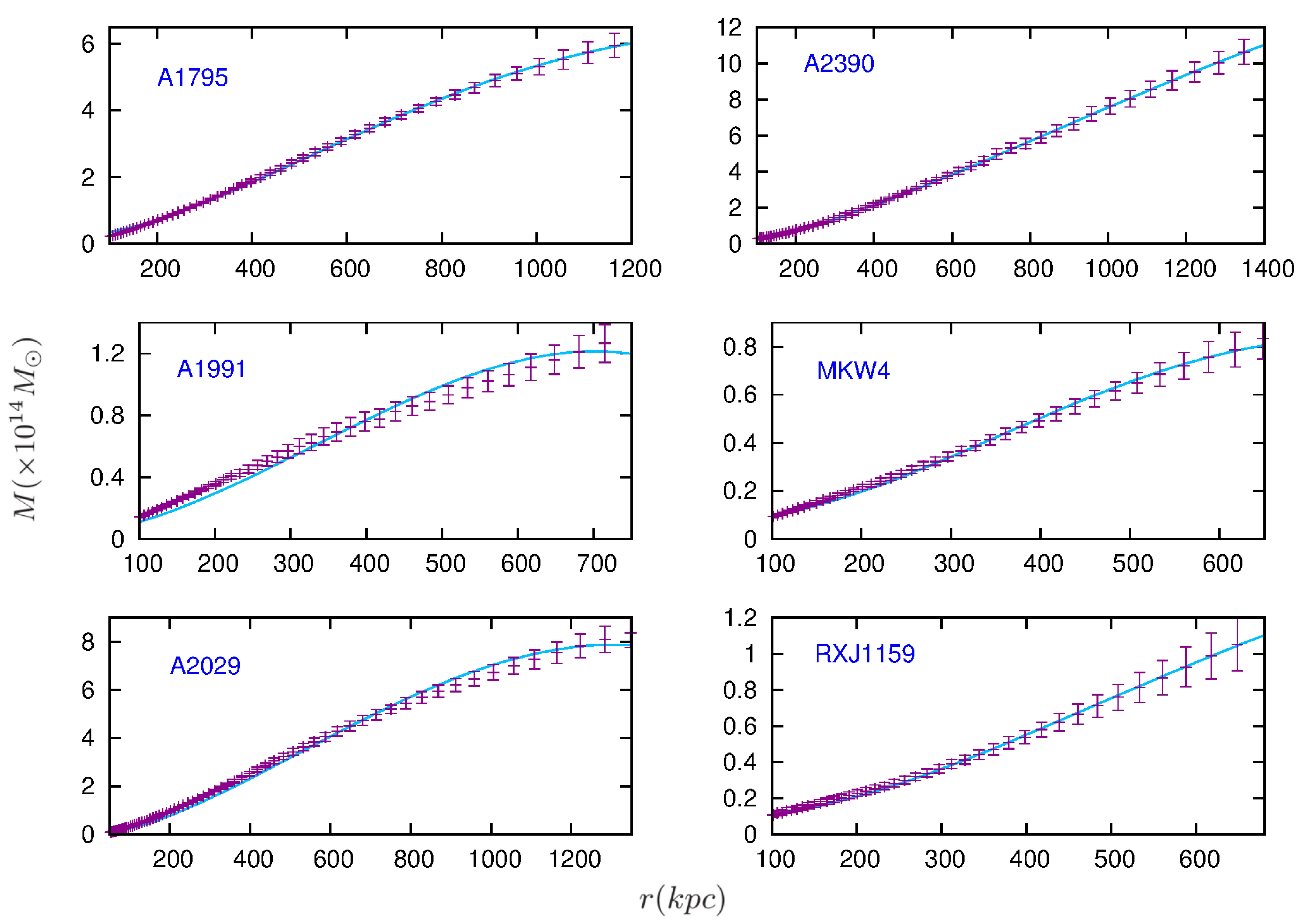}
\begin{center}  {The same as Fig.~\ref{plot_data1-6} for the left 6 clusters of galaxies.}
\end{center}
\label{plot_data7-12}
\end{figure*}

  The results for the best fits as explained in Section~\ref{cluster-profiles} are
summarized in Table~\ref{table02}.  Figs.~\ref{plot_data1-6} and~\ref{plot_data7-12} show the best fits of the theoretical masses
compared to the total dynamical ones obtained in \citet{vikhlinin06}. In all cases, the parameter $\Delta_\mathrm{AIC}<2$ and so, in general, the $f(\chi)=\chi^{3/2}$ model is good in fitting the observational data.

\begin{table*}
\small
\centering
\caption{Parameters estimation for the galaxy clusters}
\begin{tabular}{ccccccccc}
\hline
Cluster & $r_\mathrm{min}$ & 
$r_\mathrm{max}$ & $M_\text{gas}$ &
$M_\mathrm{th}$ &
$M_\mathrm{dyn}$ &
$A$ & $B$ &
$r_{s}$ \\
 & (kpc) & (kpc) & $(10^{13} \mathrm{M}_\odot)$ & $(10^{14} \mathrm{M}_\odot)$
 & $(10^{14} \mathrm{M}_\odot)$ & $(10^8 \mathrm{kpc})$ & $(\mathrm{kpc}^{-1})$
 & $(10^{-8} \mathrm{kpc})$ \\
\hline
A133 & 92.10  & 1005.81 & 3.193 & 3.269 & 3.359  & 2.0727 & 96.563 & 3.42 \\
A262 & 62.33  & 648.36 & 1.141 & 0.825 & 0.8645 & 1.7825 & 358.82 & 2.40 \\
A383 & 51.28 & 957.92 & 4.406 & 2.966 & 3.17  & 1.9714 & 151.99 & 2.96 \\
A478 & 62.33  & 1347.89 & 10.501 & 7.665 & 8.18   & 1.9271 & 61.436 & 3.91 \\
A907 & 62.33  & 1108.91 & 6.530 & 4.499 & 4.872   & 1.9024 & 101.05 & 3.56 \\
A1413 & 40.18  & 1347.89 & 9.606 & 7.915 & 8.155  & 1.9423 & 51.150 & 71293 \\
A1795 & 92.10  & 1222.57 & 6.980 & 6.071 & 6.159  & 1.9255 & 61.757 & 3.79 \\
A1991 & 40.18  & 750.55 & 1.582 & 1.198 & 1.324  & 2.3703 &  340.56 & 2.49 \\
A2029 & 31.48  & 1347.89 & 10.985 & 7.872 & 8.384  & 2.1837 & 75.339  & 440.9 \\
A2390 & 92.10  & 1415.28 & 16.621 & 11.151 & 11.21 & 1.1517 & 16.935 &  4.55 \\
MKW4 & 72.16  & 648.36 & 0.676 & 0.805 & 0.8338  & 2.4413 & 367.55 & 2.28 \\
RXJ1159 & & & & & & & & \\
+5531 & 72.16 & 680.77 & 0.753 & 1.105 & 1.119  & 2.3460 & 171.94 & 2.39 \\
\hline
 &  &  &  & & Mean value & 2.0014 & 154.59  & 5980 \\
 &  &  &  & & $<\mathrm{SD}>$ & 0.00868 & 1.6462  & 937.0 \\
\hline
\end{tabular}
\normalsize
\begin{center}
From left to right, the columns represent the name of the cluster, the minimal
$ r_\mathrm{min} $ and maximal $ r_\mathrm{max} $ radii for the integration,
the mass of the gas $ M_\mathrm{gas} $, the total theoretical mass
$ M_\mathrm{th} $ derived from our model, the total dynamical mass
$ M_\mathrm{dyn} $ from \citet{vikhlinin06}, the best-fit parameters $A$, $B$
and $r_s$, respectively. Also, at the bottom of the Table, we show the
best-fit parameters obtained from the 12 clusters of galaxies data taken as a
set of independent objective functions together, with their corresponding mean
standard deviations $<\mathrm{SD}>$.
\end{center}
\label{table02}
\end{table*}

  From the best-fit analysis, we see that our model is capable to account for
the total dynamical masses of the 12 clusters of galaxies, except at the very
inner regions for some of them, a persistent behavior more accentuated for
A907 and A1991.  Notice that the parameter $A$ quantifies the extra Newtonian-like contribution to the dynamical mass (cf.~equation~\eqref{final-acc}), and the parameters $B$ and $r_s$ are present in the $\Phi_c^{(2)}$ term only (equation~\eqref{Phi_c1}). As can be seen in Table~\ref{table02}, there are
two systems, A1413 and A2029, for which the estimated parameter $r_s$ is very
far from the mean value for the other clusters.  Comparing its contribution to the acceleration with
respect to the other two terms in equation~\eqref{final-acc}, we found that
the dominant second order term is the one with the parameter $A$, and the
contribution of the derivative of the integral~\eqref{Phi_c1} is very small
(since $r_s$ appears inside a logarithm and because of the particular combination of
the functions in such equation).

  From the figures, we see that the ``MOND-like'' relativistic correction of our
model is better at the outer regions of the galaxy clusters than standard MOND, that needs extra matter to fit the observations in these systems.  Also, the second order
perturbation analysis of the metric theory $f(\chi)=\chi^{3/2}$ was
capable to account for the observations of the rotation curves of spiral
galaxies and the Tully-Fisher relation, and the gravitational lensing in
individual, groups and clusters of galaxies \citep{mendoza13}. In this work,
we keep fixed those parameters at $\mathcal{O}(2)$ of perturbations to
obtain the $\mathcal{O}(4)$ of the model, with the additional result that it
is possible to fit the dynamical masses of clusters of galaxies without the need
of extra DM.

  Up to now it has generally been thought that a MOND-like extended theory of
gravity was not able to explain the dynamics of clusters of galaxies without
the necessary introduction of some sort of unknown DM component.  Our
aim has been to show that in order to account for this dynamical description
without the inclusion of DM, it is necessary to introduce
relativistic corrections in the proposed extended theory.  To do so, we have
chosen the particular $f(\chi) = \chi^{3/2}$ MOND-like metric extension \citep{bernal11a}, which has also shown to be in good agreement with
gravitational lensing of individual, groups and clusters of galaxies and with
the dynamics of the Universe providing an accelerated expansion without the
introduction of any dark matter and/or energy entities \citep[see][for a review]{mendoza15}. 

  A similar analogy occurred
when studying the orbit of Mercury about a century ago.  Its motions were
mostly understood with Newton's theory of gravity. However it was necessary to
add relativistic corrections to the underlying gravitational theory to account
for the precession of its orbit.  Mercury orbits at a velocity $ v \sim 50 \,
\mathrm{km}/\mathrm{s} $, implying $ v/c \sim 10^{-4} $ and
already relativistic corrections are required.  Typical velocities of clusters
of galaxies are $ v \sim 10^{3} \, \mathrm{km}/\mathrm{s} $, with
$ v/c \sim 10^{-3} $.  This means that the dynamics of clusters of galaxies are
about one order of magnitude more relativistic than the orbital velocity of
Mercury and so, if the latter required relativistic corrections, then the
necessity to describe the dynamics of clusters of galaxies with relativistic
corrections are even more important.

\section*{Acknowledgments}
We gratefully acknowledge Alexey Vikhlinin for kindly providing the data of the galaxy clusters used in this article. We thank J.C. Hidalgo for his valuable comments on the first discussions on this work, K. MacLeod for useful corrections on a previous version of this article and the anonymous referee who helped us with valuable comments to improve this article.  This work was supported by Direcci\'on General de Asuntos del Personal Acad\'emico (DGAPA)-UNAM(IN111513 and IN112019) and Consejo Nacional de Ciencia y Tecnolog\'ia (CONACyT) M\'exico (CB-2014-01 No.~240512) grants.  
TB, OLC and SM acknowledge economic support from CONACyT
(64634, 62929 and 26344).

\bibliographystyle{rmaa}
\bibliography{clusters-fchi}

\end{document}